\documentclass[citeauthoryear]{llncs}
\pdfoutput=1
\usepackage{url}
\usepackage{cite}
\usepackage{listings}
\usepackage[utf8]{inputenc}
\usepackage{graphicx}
\usepackage{float}
\usepackage{verbatim}
\usepackage{eurosym}

\title{Modeling Languages: metrics and assessing tools}

\author{Daniela Fonte \and Ismael Vilas Boas \and José Azevedo \and José João Peixoto \and Pedro Faria \and Pedro Silva \and Tiago Sá \and  Ulisses Costa \and Daniela da Cruz \and Pedro Rangel Henriques}

\institute{Department of Informatics, University of Minho\\ Campus de Gualtar, 4710-057 Braga, Portugal\\
\email{\{danielamoraisfonte, ismael.vb, jazevedo, jj.peixotopereira, pedro.faria.80, zepedro.cs, ulissesmonhecosta, danieladacruz, pedrorangelhenriques\}@gmail.com, tiago@esterisco.com}
}

\newcommand{\uml}{\textsf{UML}}
\newcommand{\xmi}{\textsf{XMI}}
\newcommand{\xml}{\textsf{XML}}
\newcommand{\entArch }{\textsf{Enterprise Architect}}
\newcommand{\sdmetrics}{\textsf{SDMetrics}}

\begin{document}
\maketitle

%%Abstract
\begin{abstract}
Any traditional engineering field has metrics to rigorously assess the quality of their products.
Engineers know that the output must satisfy the requirements, must comply with the production and market rules, and must be competitive.

Professionals in the new field of software engineering started a few years ago to define metrics to appraise their product: individual programs and software systems.
This concern motivates the need to assess not only the outcome but also the process and tools employed in its development.
In this context, assessing the quality of programming languages is a legitimate objective;
in a similar way, it makes sense to be concerned with models and modeling approaches, as more and more people start the software development process by a modeling phase.

In this paper we introduce and motivate the assessment of models quality in the Software Development cycle.
After the general discussion of this topic, we focus the attention on the most popular modeling language -- the UML -- presenting metrics.
Through a Case-Study, we present and explore two tools.
To conclude we identify what is still lacking in the tools side.

%\textbf{Note for the PCommittee: at moment the paper is 2 pages longer than
%the allowed; if accepted for publication,  we assume the obligation to
%shortener it to the appropriated size without loosing any content}

%In the paper we discuss the quality of modeling languages, introducing and motivating the topic, presenting metrics, and comparing tools.
\end{abstract}


%%Introdução
\section{Introduction}

Models are a representation of reality aiming at the simplification of some complex objects: they are built so that we can better understand the system being developed.
%They help us to visualize a system as it is or as we need it to be, allowing to specify the structure and behavior of it.
They allow us to specify the structure and behavior of a system, providing the guidance lines/blueprints for constructing a system, and finally, they document the decisions taken for a given system.
Specifying means building models that are precise, unambiguous and complete. 

Some models are best described textually, other graphically. All interesting systems exhibit structures that transcend what can be represented in a programming language.

A modeling language is a language whose vocabulary and rules focus on the conceptual and physical representation of a system. %%todo
%% prh idea
One the one hand, one can produce a strict formal specification of the system, which allows us to reason over the system proprieties, without running the system.
On the other hand, one can follow a pragmatic approach, using a diagrammatic specification of the system, not allowing us to reason over programs,
but deriving programs from the model specification. That aside, when assessing a modeling language we might need to infer on its quality.

\begin{comment}
\begin{quotation}
Effective management of any process requires quantification, measurement, and modeling.
Software metrics provide a quantitative basis for the development and validation of models of the software development process.
Metrics can be used to improve software productivity and quality\cite{g1:Millis:1998}.
\end{quotation}
\end{comment}

The main goal of using model/software metrics is to be able to generate quantifiable measurements from the specifications/software. 
According to~\cite{g1:Millis:1998}, metrics can be used to improve software productivity and quality.
The use of model metrics is even more important to numerous valuable applications in earlier stages of the development process like scheduling, cost estimation, quality assurance, and personnel task assignments.
Nowadays, this metrics become increasingly essential for Software Engineering: they are crucial even for reengineering processes.
In \emph{Forward Engineering} they are used to measure the software quality and estimate cost and effort of software projects\cite{Fenton}.
In the field of \emph{Software Evolution}, they can be both used to identify stable or unstable parts of a system as to determine where refactoring can be or have been applied\cite{Serge}.
They even can be used for assessing the quality and complexity of software systems in \emph{Software Reengineering} or \emph{Reverse Engineering}\cite{43044}.

%The \uml\ addresses the specification of all the important analysis, design and implementation decision. %todo
When focusing on the field of Object-Oriented (OO) systems, many metrics have been proposed for assessing the design of a software system.
However, most of the existing approaches involve the analysis of the source code and cannot be applied in earlier stages of the development process.
In fact, it is not always simple to apply the existing metrics in this earlier stages. 
As the \textsf{Unified Modeling Language}, proposed by Booch, Jacobson and Rumbaugh\cite{USDPuml} has became a standard for expressing, design and specify OO systems, applying metrics to these models enables an early estimate of development efforts, implementation time, complexity and cost of the system under development. \\

In this paper, we introduce and discuss the major existing metrics for \uml\ models and present a set of tools designed for measuring \uml\ projects.
In Section~\ref{metrics} we describe the principal measurements applicable to the most popular \uml\ diagrams.
Then, in Section~\ref{tools} we present two of the best tools designed to extract metrics from \uml\ models and the results of applying them to a real case-study.
Then, Section~\ref{assess} is devoted to the metrics assessment problem.
We conclude in Section~\ref{conc} with a final balance of these topics.

%%Metrics intro
\section{Applying Metrics To \uml\ Models}\label{metrics}

An \uml\ model can be made from different diagrams, each one with a distinct view of the system.
We have, in one hand, Use Cases diagrams which expose the system functional requirements and how each user interacts with them. They are a good overview of what features the system offers to the end user.
In the other hand, we use Class Diagrams for represent the blueprint of the application under the developer perspective: they illustrate which programming components a system has and how they related to each other.
Package Diagrams describe how to group the classes and how these groups relate to each other (\emph{package import}, \emph{package merge}).
%We believe that an \uml\ diagram need to have at least these three diagrams implemented, because they give to both the costumer and the developer the full knowledge of the system.
Here we present some metrics related to this three fundamental diagrams and conclude the section by introducing some metrics for other not less important \uml\ diagrams.

%%ck metrics
\subsection{Object-Oriented Software: \textrm{CK} Metrics}
One of the most popular suites of OO metrics was proposed by Chidamber and Kemerer \cite{Chidamber:1994:MSO:630808.631131} to capture different aspects of OO designs, including complexity, coupling and cohesion. 
As we can see in~\cite{Power2}, they were posteriorly adapted for modeling languages and can be easly applied to \uml\ class diagrams.

This suite is composed by six metrics: \emph{Weighted Methods Per Class} (WMC), \emph{Depth of Inheritance Tree} (DIT), \emph{Number of Children} (NOC), \emph{Response For a Class} (RFC), \emph{Coupling between Object Classes} (CBO) and, finally, \emph{Lack of Cohesion in Methods} (LCOM).
We detail bellow each metric and its features.

\paragraph{Weighted methods per class} (WMC) - This metric regards to the complexity of a class method, being equal to the sum of those methods complexities. There are two kinds of WMC metrics:
\begin{itemize}
\item \textbf{$WMC1_{1}$} is computed from a class diagram by counting the number of methods in that class - considering each method as an unity;
\item \textbf{$WMC_{cc}$} is computed by counting the number of methods in each class, based on the result of the McCabe Cyclomatic Complexity of each method.
%This metric requires information from other kind of diagrams, like Sequence or Activity diagrams.
\end{itemize}

\paragraph{Depth of inheritance tree} (DIT) - This metric is equal to the maximum length from the class to the root of the inheritance, which could be defined as the depth of the class. It is computed by taking the union of all the class diagrams in a \uml\ model and traversing the inheritance hierarchy of the class.

\paragraph{Number of children} (NOC) - This metric represents the number of childs and descendants of a certain class. Can be obtained gathering all diagrams class, in a \uml\ modulation, and checking all the inheritance relations of the class.

\paragraph{Response for a class} (RFC) - This metric measures the number of methods that can be invoked by an object of a given class. It can be obtained from a class diagram and from behavior diagrams (e.g. sequence diagrams), which can inform of several methods of other classes that are invoked by each of the class methods.

\paragraph{Coupling between object classes} (CBO) - Two classes are related if a method of a class uses a instance variable or method of another class.
Thus, we can compute this metric by counting the number of classes to which the class is related and counting all kind of references of the attributes and parameters of the class methods.
%Counting the number of classes to which the class is related and counting all kind of references of the attributes and parameters of the class methods, an estimate of this metric value can be obtained from the class diagrams. 
Though, it is possible to calculate a more precise value if behavioral diagrams are taken into account, since the usage of instance variable and invocation methods are additional information.

\paragraph{Lack of cohesion in methods} (LCOM) - It measures the number of sets of instance variables accessed by every pair of methods of a given class, that has a non-empty intersection. For this, is essential to use the information of the usage instance variables by the methods of a class -- i.e., since a class diagram does not have information about the usage, it is required a sequence diagram.\\

This set of metrics can be found and cited in several papers -- like~\cite{Power2} -- and represent the basis of all the existing metrics for OO systems.

%%Diagram class metrics
\subsection{Class Diagram and Package Metrics}

These diagrams are used to describe the types of objects in a system and the relationships among them.
They describe the structure of a system by showing the system classes, their attributes and methods or operations.
Their quality have a huge impact on the final quality of the software under development, as they describe the general model of the system information.

\begin{table}
\begin{minipage}[b]{0.5\linewidth}\centering
\begin{tabular}{ p{1,5cm} | p{4cm}}
\multicolumn{2}{l}{\textbf{Marchesi Metrics}} \\ \hline
\textbf{Metric} & \textbf{Description} \\ \hline
NC & Number of Classes \\ \hline
CL1 & Weighted Number of Class responsibilities   \\ \hline 
CL2 & Weighted Number of Class Dependencies \\ \hline 
CL3 & Depth of inheritance tree \\ \hline 
CL4 & Number of immediate subclasses of a given class \\ \hline 
CL5 & Number of distinct class \\ \hline 
\end{tabular}
\caption{\small{Marchesi Class Diagram Metrics}}
\label{t:dcm}
%\end{table}
%\end{center}
\end{minipage}
\hspace{0.3cm}
\begin{minipage}[b]{0.5\linewidth}
\centering
%\begin{center}
%\begin{table}[h]
\begin{tabular}{ p{1,5cm} | p{4cm}}
\multicolumn{2}{l}{\textbf{Marchesi Metrics}} \\ \hline
\textbf{Metric} & \textbf{Description} \\ \hline
NP & Number of Packages. \\ \hline 
PK1 & Number of Classes \\ \hline
PK2 & Weighted Number of responsibilities of a Class   \\ \hline 
PK3 & Overall Coupling among Packages \\ \hline 
\end{tabular}
\caption{\small{Marchesi Package Metrics}}
\label{t:pcm}
%\end{table}
%\end{center}
\vspace{0.78cm}
\end{minipage}
\end{table}

Measures like the \emph{Number of Attributes in the Class}, the \emph{Number of Operations in the Class}, \emph{Number of Inherited Attributes}, \emph{Number of descents/ancestors of a Class}, or even the \emph{Number of Interfaces Implemented} are used both for indicate the system complexity as for an index of quality.
Many works present several metrics for this diagrams~\cite{DBLP:journals/Lobjet/GeneroPC00},~\cite{Eichelberger_onclass},~\cite{Yi04acomparison}.
The OOA metrics defined in~\cite{Marchesi:1998:OMU:522081.795010} also contemplate Class Diagrams, as we can see in Table~\ref{t:dcm}.

In \uml, classes can be grouped into Packages to define subsystems or even for implementation purposes.
The measurement of a package complexity is useful to forecast the development efforts of it.
For that, we can measure properties like the \emph{Number of Classes of a Package}, the \emph{Total Number of Packages in the system}, or the \emph{Number of Interfaces in the Package}.
Marchesi~\cite{Marchesi:1998:OMU:522081.795010} suggests several Package Metrics as described in Table~\ref{t:pcm} for estimate this complexity.

%%Use Case metrics
\subsection{Use Case Metrics}

Use Cases Diagrams are graphical representations of entities which interact with the system (\emph{actors}) and operations that the system must perform for them.
They define a sequence of actions which illustrate a specific way of using the system.

These diagrams are functional specifications, collected at the beginning of a system development process.
They are crucial to an early estimate of the system complexity and its development efforts, as we can see by the UC metrics defined in several works like~\cite{Kim02developingsoftware},\cite{Mohagheghi05effortestimation}, \cite{Ribu01estimatingobject-oriented}.

In fact, measuring the number of Use Cases, actors and communications among them is a good indicator of the system complexity, as well as to quantify the relationship between diagrams (i.e. estimates the number of UC that extend or include others).

One remarkable work on this area was performed by Michele Marchesi~\cite{Marchesi:1998:OMU:522081.795010}.
Table~\ref{t:ucm} illustrates the Use Case metrics defined on this work. 
 
\begin{table}[h]\centering
\begin{tabular}{ p{1,5cm} | p{10.5cm}}
\multicolumn{2}{l}{\textbf{Marchesi Metrics}} \\ \hline
\textbf{Metric} & \textbf{Description} \\ \hline
NA & Number of actors of the system. \\ \hline
UC1 & Number of Use Cases in the system. \\ \hline 
UC2 & Number of communications among UC and Actors  \\ \hline 
UC3 & Number of communications among UC and Actores without redundancies \\ \hline 
UC4 & Global complexity of the system \\ \hline 
\end{tabular}
\caption{\small{Marchesi Use Case Metrics}}
\label{t:ucm}
\end{table}

The \textbf{UC4} metric represents a balance of the global complexity of the system, and its value is obtained through the values of \textbf{UC1}, \textbf{UC2} and \textbf{UC3} metrics.

%%Other
\subsection{Other Diagram Metrics}

Statechart diagrams illustrate the behavior of an object.
They define different states of an object during its lifetime, which are changed by events.
A \emph{state} expresses an action of an object during a certain time, when it does not receive external stimulus nor is there any change in its attributes. 

Measures like the \emph{Number of Entry Actions}, \emph{Number of Exit Actions}, \emph{Number of Transitions}, or even the \emph{Number of Activities} are associated to the complexity and dimension of the problem~\cite{EVMmdm}.
In Table~\ref{t:estado} we can notice some examples from measurable attributes for this type of diagram.

\begin{table}
\begin{minipage}[b]{0.5\linewidth}
\centering
\begin{tabular}{ p{1,5cm} | p{4cm}}

\multicolumn{2}{l}{\textbf{Statechart Metrics}} \\ \hline
\textbf{Metric}  & \textbf{Description} \\ \hline
TEffects  & Number of transitions with an effect in the state machine. \\ \hline 
TGuard & Number of transitions with a guard in the state machine. \\ \hline 
TTrigger & Number of triggers of the state machine transitions. \\ \hline 
States & Number of states in the state machine. \\ \hline 
\end{tabular}
\caption{\small{Statechart Diagrams Example}}
\label{t:estado}
%\end{table}

\end{minipage}
\hspace{0.3cm}
\begin{minipage}[b]{0.5\linewidth}
\centering

\begin{tabular}{ p{1,5cm} | p{4cm}}
\multicolumn{2}{l}{\textbf{Activity Metrics}} \\ \hline
\textbf{Metric} & \textbf{Description} \\ \hline
Actions  & Number of activity actions. \\ \hline 
Object-Nodes & Number of activity object nodes. \\ \hline 
Pins  & Number of pins on the activity nodes. \\ \hline 
Guards  & Number of guards defined on object and control flows of the activity. \\ \hline 
\end{tabular}
\caption{\small{Example of Activity Diagrams}}
\label{t:act}
\vspace{0.25cm}
\end{minipage}
\end{table}

Activity diagrams describe work flows and are very useful for detail operations of a class (including behaviors expressed by parallel processing).
As we can see in Table~\ref{t:act} several metrics for this diagrams are available.

Besides these metrics, it is possible to measure attributes like the \emph{Number of Activity Groups/Zones}, the \emph{Number of object flows} or even the \emph{Number of Exceptions} of each diagram.

After this methodological research through which we introduce the more consistent and relevant metrics found in this area, we present in the next section tools for apply them to \uml\ models and put them to test with a real case-study.

%%Tools
\section{Tools for \uml\ Metrics Calculation} \label{tools}

Nowadays, it is very common to use tools like \textsf{Visual Paradigm for \uml\footnote{Available at \url{http://www.visual-paradigm.com/product/vpuml/}}} or even \textsf{Poseidon for \uml}\footnote{Available at \url{http://www.gentleware.com/products.html}} for software application development.
They offer a visual environment to model software, which reduces the complexity of software design.
However, they do not support metrics specification - it is necessary to use other tools, designed for this task.
In the next subsections, we will introduce two systems for quantitative analysis of the structural properties of \uml\ models, and put them to test for exploring their features with a real case-study.
% (defined bellow).

One of the tools that we are going to address is \sdmetrics\footnote{Available at \url{http://www.sdmetrics.com/}}, a design measurement tool for \uml\ models.
Although its core is open source and available under the GNU Affero General Public License, \sdmetrics\ GUI it is not freely distributed. 
%The core functionalities of this system include:
%\begin{itemize}
%\item the configurable \xmi\ parser for \xmi1.0/1.1/1.2/2.0/2.1 input files;
%\item the metrics engine to calculate the user-defined design metrics;
%\item the rule engine to check the user-defined design rules.
%\end{itemize}
It is a very complete design measurement tool, analyzing a wide range of \uml\ diagrams, including Class, Use Case, Activity and Statemachine diagrams, generating several metrics for each type of diagram.

The other tool we test is the \textsf{Sparx System Enterprise Architect}{\footnote{Available at \url{http://www.sparxsystems.com.au}}}, a team-based modeling environment. 
It embraces the full product development lifecycle, supporting both software design, requirements management, and metrics calculation for Use Case Diagrams.
It allows to estimate the complexity of the project in an earlier stage, as well as the complexity associated with each system actor.

%%Case Study
\subsection*{Case Study}
Our case-study is the model built to describe an information system for the \textit{MWK (Manages With Knowledge)} service management company\footnote{This modeling project was developed in the context of the Software Systems Development master course}.

\begin{figure}[!htbp]
\begin{center}
\includegraphics[scale=0.345]{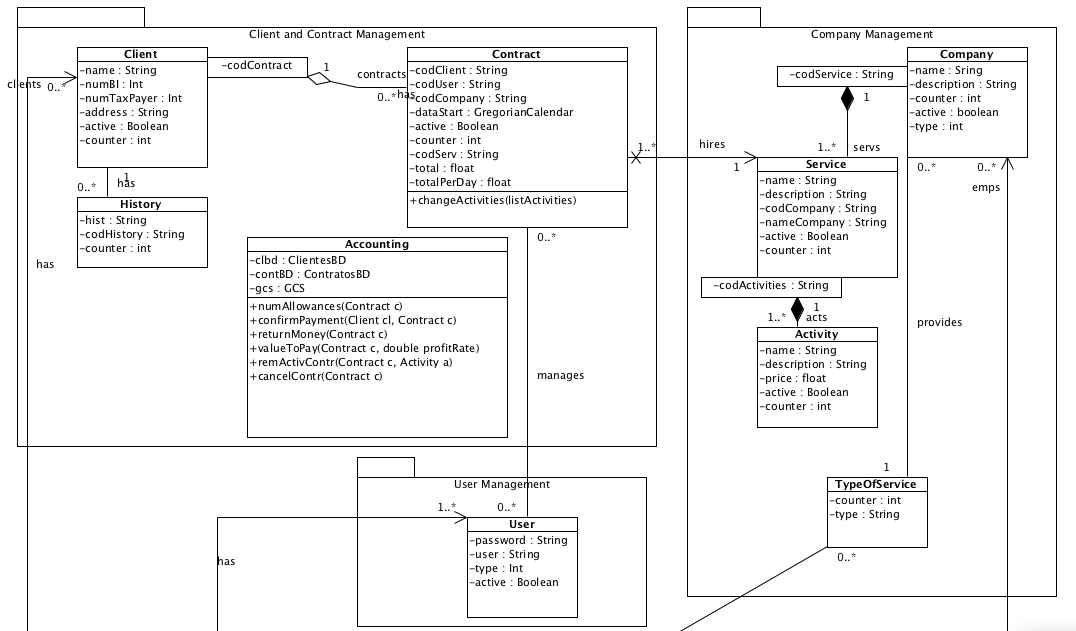}
\caption{Excerpt of a Class diagram}\label{fig:class}
\end{center}
\end{figure} 

In order to meet its clients needs \emph{MKW} is a company that has a wide range of suppliers subcontracted to be responsible for services execution.
Multiple suppliers can supply the same service and each service can be delivered in different ways.
Each service can then be composed of multiple activities. As an example, there could be a service called \textit{Shirts until 10 Kg} and inside this service there could be activities such as \textit{wash, iron, sewing buttons, etc}.
Each activity of a given service as a stipulated price, and can be hired by a client.
%The goal of the project was to model and implement a management software for this task.

\begin{figure}[!htbp]
\begin{center}
\includegraphics[width=0.9\textwidth]{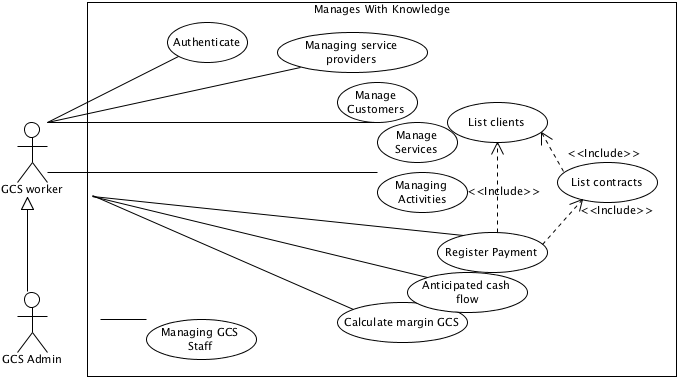}
\caption{The general Use Case model}\label{fig:usecase}
\end{center}
\end{figure} 

The complete \emph{MKW} model is composed of Use Case, Class, Sequence and Statemachine diagrams.
As an example of the \uml\ diagrams used to model this task, we can see in Figure \ref{fig:usecase} an image of a general Use Case diagram and, in Figure \ref{fig:class}, an excerpt of a Class diagram.

%Adding to Use Case and Class diagrams, the modelation is composed of Sequence and State Machine diagrams.

%SDMetrics
\subsection{SDMetrics}
\sdmetrics\ is a design measurement tool for analyze a wide range of \uml\ diagrams, including Class, Use Case, Activity and Statemachine diagrams.

For each type of diagram, this tool generates several metrics.
For example, the \textbf{NumAttr} metric is calculated from Class diagrams and measures the number of attributes in a class.
Other one is \textbf{ExtPts}, which is calculated from Use Case diagrams, and gives us the number of extension points of a given use case.

\sdmetrics\ is written in \textsf{Java}, and provides us a graphical user interface for analyze \xmi\footnote{
\xmi\ (\textbf{X}ML \textbf{M}etadata \textbf{I}nterchange)  is an \emph{OMG} (Object Management Group) standard to generate \xml-based representations of \uml\ and other OO data models.} 
files, most modeling tools support project exportation in \xmi.

This tool allows to access the results from different views. We will introduce the ones that seem the most important:
\begin{itemize}

\item \textbf{Metric Data Tables} provides a table that matches each \uml\ model element analyzed (table line) to its value for each metric (table column);
\item \textbf{Histograms} provides a graphical distribution  for each design metric;
\item \textbf{Design Comparison} provides us a mean to compare the structural properties of two \uml\ designs. It is very useful to compare the same design in different iterations of the development, or to compare an alternative design to the current one.
\item \textbf{Rule Checker} design rules and heuristics detect potential problems in the \uml\ design such as incomplete design (i.e. unnamed classes, states without transitions, etc.);  violation of naming conventions for classes, attributes, operations, packages; etc;

\begin{comment}
	\begin{itemize}
	\item incomplete design such as unnamed classes, states without transitions;
	\item violation of naming conventions for classes, attributes, operations, packages;
	\item etc.
	\end{itemize}
\end{comment}

\item \textbf{Catalog} this view provides us with the definitions of the metrics, design rules, and relation matrices for the current data set, and provides literature references and a glossary for them.
\end{itemize}

One of the most advanced features in this software is the possibility of defining Custom Design Metrics and Rules. 
The new metrics are defined in a \xml file, with a very particular format, the \emph{SDMetricsML} (\sdmetrics Markup Language).

The \sdmetrics\ tool does not provide a direct notion of good or bad quality of the design model. 
Despite that, on its User Manual\footnote{\url{http://www.sdmetrics.com/manual/index.html}} we can find tips of how to interpret each kind of metrics. 
 
\subsubsection{Results}
\begin{figure}[htbp]
\includegraphics[scale=0.3]{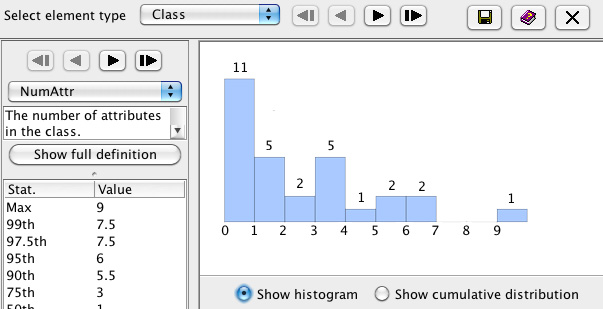}
\hspace{0.1cm}
\includegraphics[scale=0.32]{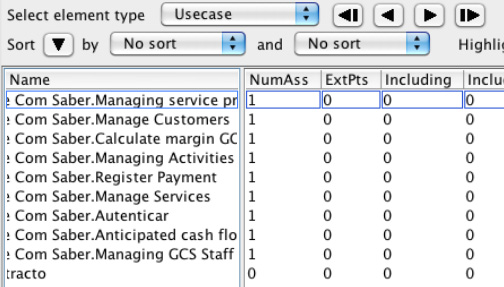}
\caption{\sdmetrics: \textsf{NumAttr} Histogram and Metrics Table for Use Case Diagrams}
\label{fig:sdmetrics}
\end{figure}

Based on the \sdmetrics\ manual, we will now explain how to analyze each metric obtained. 
Figure~\ref{fig:sdmetrics} illustrates some outputs of \sdmetrics\ for our case-study. 
On the left, we can see an Histogram for class diagrams evaluating the \emph{NumAttr} metric. On the right, we present an excerpt of a general metrics table for class diagrams.
%Next we will show some printscreen taken from SDMetrics, which will illustrate some of the outputs of this software, for our case-study.\\

\begin{comment}
\textbf{Metrics Table}
\begin{figure}[H]
\begin{center}
\includegraphics[width=1\textwidth]{images/table.png}
\caption{Metrics Table for class diagrams}\label{img:table}
\end{center}
\end{figure} 

\textbf{Histogram}
\begin{figure}[H]
\begin{center}
\includegraphics[width=1\textwidth]{images/histogram.png}
\caption{Histogram for class diagrams evaluating the metric NumAttr}\label{img:histogram}
\end{center}
\end{figure} 

\textbf{Rule Checker}
\begin{figure}[H]
\begin{center}
\includegraphics[width=1\textwidth]{images/rule.png}
\caption{Rule checking for Operations}\label{img:rule}
\end{center}
\end{figure} 
\end{comment}

Size metrics usually count elements inside design elements (class, package, etc).
They are good to estimate developing costs and effort, and can be directly found on the Metrics Table.
A large design element may indicate that it suffers from poor design, resulting in low functional cohesion. This has a negative impact on the understandability, reusability, and maintainability of the design element.

%The coupling between design elements is also measured by \sdmetrics. 
Measure \emph{Coupling} is estimate the degree of elements connection in a design: the more they are connected, the more they depend on each others. 
%Coupling is the degree to which the elements in a design are connected: the more they are connected, the more they depend on each others. 
Changing a design element may force the user to adapt the connected elements; also a problem in a design element may cause failure in a completely different connected element. 
An high degree of dependency may affect the system maintainability and testability -- it is crucial to minimize coupling.

Inheritance-related metrics usually calculate features such as depth/width of the inheritance and number of ancestors/descendants of a design element. Such as high coupled elements, deep inheritance structures are believed to be more fault-prone. It is harder to fully understand a class that is situated deep in the inheritance tree, because you have to understand its ancestors. Also, modifying a design element with many descendants, may have a large impact on the system.

Complexity metrics measure the degree of connectivity between elements of a design element. They are concerned with relationships/dependencies between the elements in the design unit, such as the number of method invocations inside a class. The high complexity between the elements of a design element can make the design harder to understand, therefore more propitious to faults. Complexity metrics are usually strongly correlated with size measures. So even though they are good indicators of quality, such as fault-proneness, they provide no new knowledge comparing to size metrics.

In general, these guidelines lead us to belief that the lower the metrics values are, the better.
In what concerns our case-study, after observed the metrics obtained we have noticed that \emph{GCS} class seems to have too many operations (26), specially compared with the rest of the classes.
On the other hand, some classes look like they are missing operations.
These conclusions suggest a careful analysis of the project to identify classes which operations do not belong to them and should be given to other classes; also classes that lack operations, should be completed.
This is just an example of how the output of \sdmetrics\ could be useful during a modeling process.

\begin{comment}
In general, this directions lead us to belief that the lower the metrics values are, the better. Looking at the modelation which is our case study, it is not easy to establish  a maximum value for each metric. Either way, and looking at the modelation as an all, \textit{GCS} class for example, seems to have to many operations (26), specially compared to the rest of the classes. On the other hand, some classes look like they are missing operations. Taking this in regard, we should analyze carefully this class, and perceive which operations do not belong to it, and should be given to other classes. Also classes that lack operations, should be completed.
This is just an example of how the output of \sdmetrics\ could be useful during the modelation process.
\end{comment}

%Sparkx Systems
\subsection{Sparx Systems Enterprise Architect}
Sparx Systems \entArch~is another tool that provides modeling of \uml\ diagrams.
It supports mind map diagrams and project management, to provide full traceability from requirements specification to deployment end implementation.
This tool also provides some metrics evaluation to compute the complexity of a project based on Use Case diagrams. 

To perform this evaluation, the user needs to provide a level of implementation complexity for each interaction with authors. 
This task can be done when defining the Use Case descriptions or when performing the metrics evaluation.

To evaluate the metrics, \entArch~has a wizard which enables other options for the complexity analysis.
These options manipulate the \emph{Technical} and the \emph{Environment} complexities and are used to adapt the evaluation and perform a better result on estimation.

This system enables filtering the Use Cases used in evaluation, both on manual selection or regular expressions over use case name.
%Also can be filtered the Use Cases used in evaluation. 
%To filter can be used manual selection or regular expressions over use case name.
This kind of filtering enables the project to be distributed and evaluated individually.

\subsubsection{Results}
For use this tool in metrics calculation, the user have the possibility to do some tweaking over use case complexity (on their description) to get more precise results.
This complexity admits simple values like low, medium or high.

\begin{figure}[!htbp]
\includegraphics[scale=0.257]{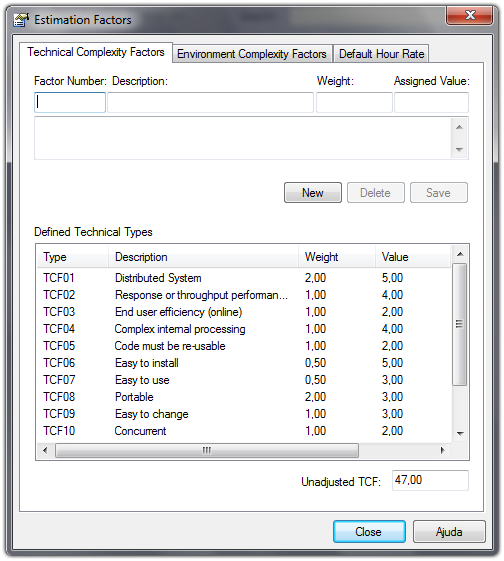}
\hspace{0.1cm}
\includegraphics[scale=0.29]{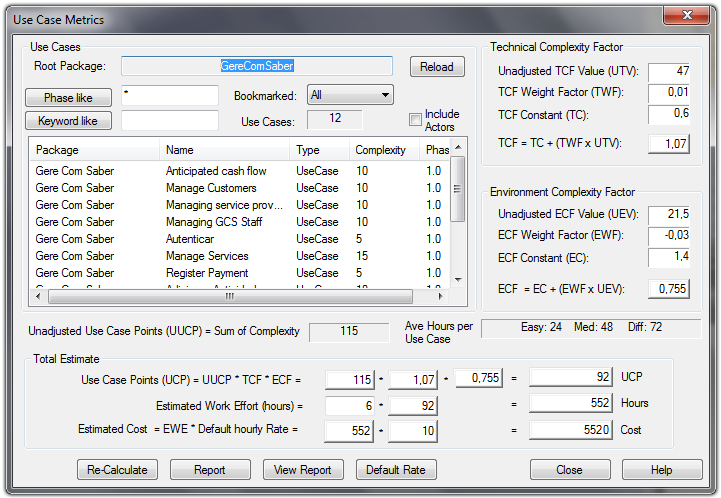}
\caption{Estimation Factors and Use Case Metrics of \entArch~wizards}\label{img:sparxRes}
\end{figure}

\entArch~offers a wizard which enables the edition of factors related to environment and technics complexity or even the hour rates, as we can see on the left side of Figure \ref{img:sparxRes}.
%Some factors related to environment and technics complexity could also be edited or about hour rates in the wizard shown on the left side of Figure \ref{img:sparxRes}.
Here we can use multiple factors to evaluate this factors, like usability or portability on technical complexity.

We can also access a metrics wizard, as depicted in the right side of Figure \ref{img:sparxRes}, which illustrates the set of default values used to evaluate the task effort.
A set of predefined values are based on the factors edited on the previous wizard.

The final result of the metrics evaluation process is an estimation of Working Hours, Use Case Points\cite{Ribu01estimatingobject-oriented} and Total Cost needed to perform the system development.
In our case-study, the project has twelve Use Cases and many of them have medium complexity. 
Thus, as we can see in the right side of Figure \ref{img:sparxRes}, the effort to complete the task is 552 working hours, that would give a final result of \EUR{5.520}. 
For obtaining this value only was changed the use cases complexity and everything else was left with default values.

%%Metrics Assessment
\section{Metrics Assessment} \label{assess}
\begin{comment}
Effective management of any process requires quantification, measurement, and modeling.
Software metrics provide a quantitative basis for the development and validation of models of the software development process.
Metrics can be used to improve software productivity and quality\cite{g1:Millis:1998}.
\end{comment}

%And so, we need to model complex sistems. 
When interpreting the values obtained from measurements, one might question: \emph{is the metric really measuring the intended attribute?} This is a question that is present in the industry, yet unsuccessfully answered. 

Working with models, one might want to know the quality of its model, i.e., which amount of it really reflects object proprieties. To discuss model quality, one must use metrics to quantify those proprieties. Fenton~\cite{g1:Fenton:1999} estimates that companies spend about 4\% of the development budget in the establishment of metrics programs, therefore, engineers should also guarantee that the applied metrics actually quantify, measure, and model the attributes of the system.

Kaner and Bond~\cite{g1:kaner:2004} proposed a framework for metric evaluation, saying that if a project or company is managed according to the results of inadequately validated metrics,and  not tightly linked to the attributes they are intended to measure, distortions and disfunctionalities should be commonplace.

\begin{comment}
	This has a likely consequence: if a project or company is managed according to the results of measurements, and those metrics are inadequately validated, insufficiently understood, and not tightly linked to the attributes they are intended to measure, measurement distortions and dysfunctional should be commonplace\cite{g1:kaner:2004}.
\end{comment}

The industry, although not having a formal answer to this question, has advanced some steps forward in this direction. The IEEE Standard 1061~\cite{g1:Ieee1061:1998} defines an \emph{attribute} as \emph{``a measurable physical or abstract property of an entity''}. A \emph{quality factor} is a type of attribute,
\emph{``a management-oriented attribute of software that contributes to its quality''}. A \emph{metric} is defined as being a {\bf measurement function}, and a {\bf software quality metric} is defined as \emph{``a function whose inputs are software data and whose output is a single numerical value that can be interpreted as the degree to which software possesses a given attribute that affects its quality''}. Any software metric must comply with the following criteria: correlation, consistency, tracking, predictability, discriminative power and reliability.
This provides a sound layout of a methodology for developing metrics for software quality attributes.

%* explicar alguns critérios\\
%* ex(a correlação dá isto, enquanto que o tracking dá-nos isto)\\

%%Conclusão
\section{Conclusion} \label{conc}

In general, the software development process follows a systematic approach aiming at a product/system of quality.
The quality criterion is not limited to the attributes of the final product, whether they comply with industry standards or not; it also ensures that the software fulfills all the specified requirements.
Most of the existing approaches that include metrics on the software lifecycle involve source code analysis and cannot be applied in earlier stages of the development process.
Applying metrics to \uml\ models enables the estimation of development effort at an earlier stage, as well as implementation time, complexity and cost of the system under development.

In this paper we focus on the most relevant metrics for \uml\ models and on two tools capable of measuring them: \sdmetrics\ and  \entArch\ systems.
We believe that our research gathers the more consistent and relevant metrics for assessing the quality of \uml\ models. 
The paper aims at offering an overview of the advantages of an early estimation of the process efforts, and to provide guidelines to the most important works on this area
Combined with a study of \uml\ metrics extraction tools, it represents an important support for an end user which needs to pick up a tool that best suits its needs.

%A qualidade na ̃o passa apenas pela obtenc ̧a ̃o de software que respeita as diversas normas e padro ̃es de qualidade, mas tambe ́m pela garantia de que este cumpre todos os requisitos especificados. As medic ̧o ̃es quantitativas associadas a esta problema ́tica revelam-se essenciais em qualquer cieˆncia, o que na ̃o e ́ excepc ̧a ̃o na Engenharia de Software: existe um esforc ̧o cont ́ınuo para encontrar novas propostas, adequadas ao desenvolvimento de software. Com este artigo pretendemos assim mostrar como tirar partido das Linguagens de Modelac ̧a ̃o existentes, e aplicar estas medic ̧o ̃es qualitativas numa fase mais inicial do projecto (aquando a sua modelac ̧a ̃o), de forma a na ̃o centrar a ana ́lise apenas no co ́digo fonte. Apresentamos assim as principais me ́tricas existentes dentro da a ́rea da medic ̧a ̃o de modelos de UML e algumas das principais ferramentas para especificac ̧a ̃o de me ́tricas sobre UML, que combinado com a aplicac ̧a ̃o de me ́tricas tambe ́m ao co ́digo fonte se revela um aliado fundamental para uma melhor gesta ̃o do software produzido.

%SDMetrics
We can conclude that \sdmetrics\ is a versatile tool to calculate a large set of metrics over a wide range of \uml\ diagrams. Based on this metrics we can try to measure the quality and complexity of a software model.
It has an interesting GUI which provides several output views, from simple \emph{tables} to \emph{Histograms}. It finds potential problems with the model and also enables new metrics definition.

The major disadvantage of this tool is the incapability of giving the user a plain and simple notion of the model quality, although \sdmetrics\ Manual provides simple tips of how to interpret each kind of metric -- crucial for a correct results reading.
At a glance, \sdmetrics\ results are guidelines for finding the good and the bad points of  \uml\ models, not a full specification of the models quality. 
%Although good documentation is provided,  which provide guidelines to the result interpretation.

On the other hand, \entArch\ is a full formal \uml\ specification environment that supports metrics calculation oriented to Use Case diagrams. 
It is driven to enterprise market and oriented to minimize the cost and time spent with the production process.
This tool provides simple results and represents a good choice to have an estimation of the
implementation costs based on a formal system specification.
With its system wizard, the user can adapt the value of the factors related to the environment and technics complexity, to obtain an accurate estimation of the system final cost.
This requires extra user interaction in the specification of a task complexity and a large knowledge of the system under development.
This complexity could be archived by other means if other types of diagrams were also analyzed.
Summing up, \entArch\ does not provide any kind of quality model analysis or do any automatic system complexity evaluation: it estimates the final cost and effort of the project.

\begin{comment}
%bad
We consider this a valid approach, but it requires some extra user interaction in the specification of the complexity of a task.
This complexity could be archived by other means if other types of diagrams were also analyzed.
%good
The final results presented are simple and also oriented to the final cost, but if the complexity has been specified with knowledge of the environment, they could estimate very well the final cost of the implemented system.

Summing up, in metrics evaluation, this tool is a good choice to have an estimation of the implementation cost on a formally specified system, but this is its only goal, it cannot provide any kind of analysis about the quality of specification or do any automatic analysis about the system complexity.
\end{comment}

\section*{Acknowledgments}
The authors would like to thank to the SDMetrics staff for provide us an Academic License to explore the full system features presented in this document.
%%Biblio
\bibliographystyle{alpha}
\bibliography{amlmt-paper}

\end{document}